\documentclass[twocolumn,english,nofootinbib,bibnotes,superscriptaddress,prb]{revtex4-1}
\usepackage{graphicx}
\usepackage[usenames,dvipsnames]{color}
\usepackage{dcolumn}
\usepackage{bbm}
\usepackage{bm}
\usepackage{amsmath}
\usepackage{amssymb}
\usepackage{times}
\usepackage{placeins}
\usepackage{hyperref}
\hypersetup{colorlinks=false}
\usepackage[version=3]{mhchem}
\usepackage{csquotes}

\graphicspath{{figures/compressed/}}

\newcommand{\nodagger}{{\phantom\dagger}}

\begin{document}

\title{Functional renormalization group approach to SU(\textit{N}) Heisenberg models:\\
	Real-space RG at arbitrary \textit{N} }
\author{Finn Lasse Buessen}
\affiliation{Institute for Theoretical Physics, University of Cologne, 50937 Cologne, Germany}
\author{Dietrich Roscher}
\affiliation{Institute for Theoretical Physics, University of Cologne, 50937 Cologne, Germany}
\affiliation{Department of Physics, Simon Fraser University, Burnaby, British Columbia, Canada V5A 1S6}
\author{Sebastian Diehl}
\affiliation{Institute for Theoretical Physics, University of Cologne, 50937 Cologne, Germany}
\author{Simon Trebst}
\affiliation{Institute for Theoretical Physics, University of Cologne, 50937 Cologne, Germany}
\date{\today}

\begin{abstract}
The pseudofermion functional renormalization group (pf-FRG) is one of the few numerical approaches that 
has been demonstrated to 
quantitatively determine
the ordering tendencies of frustrated quantum magnets in two and three spatial dimensions. 
The approach, however, relies on a number of presumptions and approximations, 
in particular the choice of pseudofermion decomposition
and the truncation of an infinite number of flow equations to a finite set.
Here we generalize the pf-FRG approach
 to SU($N$)-spin systems with arbitrary $N$ and demonstrate that the scheme becomes 
exact in the large-$N$ limit. Numerically solving the generalized real-space renormalization group equations for arbitrary $N$,
we can make a stringent connection between the physically most significant case of SU(2)-spins and more accessible SU($N$) models. 
In a case study of the square-lattice SU($N$) Heisenberg antiferromagnet, we explicitly demonstrate that the generalized pf-FRG approach is capable of identifying the instability indicating the transition into a staggered flux spin liquid ground state in these models for large, but finite values of $N$. 
In a companion paper \cite{Roscher2017} we formulate a momentum-space pf-FRG approach for SU($N$) spin models
that allows us to explicitly study the large-$N$ limit and access the low-temperature spin liquid phase. 

\end{abstract}

\maketitle


\section{Introduction}
\label{sec:intro}

Frustrated quantum magnets give rise to one of the most fascinating quantum many-body phenomena -- the formation of quantum spin liquids \cite{Anderson1973}. These highly unorthodox quantum ground states can exhibit macroscopic entanglement, while their fundamental excitations carry fractional quantum numbers \cite{Balents2010,Savary2017}. 
The latter are not only manifestly distinct from the constituent spin degrees of freedom, but in fact reveal the emergence
of a much larger underlying structure -- a lattice gauge theory in its deconfined regime.
While spin liquids are conceptually well understood by now, their unambiguous identification in microscopic model systems has remained one of the grand challenges in the field of quantum magnetism. Analytically, progress has been driven by the seminal  work of Kitaev on two paradigmatic spin models -- the toric code \cite{Kitaev2003} and the honeycomb Kitaev model \cite{Kitaev2006} -- that are both amenable to a rigorous analytical solution revealing Abelian and non-Abelian fractional excitations and an underlying Z$_2$ gauge structure
\footnote{Noticeably, Kitaev's work on the honeycomb model can readily be generalized to three-dimensional settings thereby allowing to also construct analytically solvable 3D spin models of a variety of Z$_2$ quantum spin liquids \cite{Mandal2009,Hermanns2014,Hermanns2015,OBrien2016,Yamada2017}.}. 
In parallel work, Wen has expanded the parton construction, originally developed for fractional quantum Hall liquids \cite{Jain1990,Wen1992}, to capture more general lattice gauge settings such as U(1) gauge theories or chiral Chern-Simons theories \cite{WenBook}. On the numerical side, frustrated quantum magnets have long remained out of reach for some of the most powerful simulation techniques \cite{Laeuchli2011}, such as quantum Monte Carlo approaches that typically suffer from the notorious sign problem in the presence of frustration.
For two-dimensional settings, significant progress has been made by pushing the development of the density matrix renormalization group (DMRG) towards the simulation of quasi two-dimensional ladder systems \cite{Stoudenmire2012}.
Three-dimensional systems, however, have remained elusive for almost all numerical approaches. 
One remarkable exception is the development of the pseudofermion functional renormalization group (pf-FRG) \cite{Reuther2010}, which has been demonstrated to identify the presence or absence of magnetic ordering tendencies in a number of three-dimensional quantum magnets \cite{Iqbal2016,Buessen2016,Iqbal2017,Buessen2017}. 
Technically, the pf-FRG approach starts from a decomposition of the spin degrees of freedom into auxiliary (pseudo)fermions, akin to a Schwinger fermion representation or Wen's parton construction, and then employs the functional renormalization group scheme introduced by Wetterich \cite{Wetterich1993} for the study of many-fermion problems \cite{SalmkampFRGBase01}. 
In the discussion of interacting spin models, the application of the pf-FRG approach has been met with some skepticism as the underlying FRG scheme involves a number of presumptions and approximations -- though a variety of undeterred numerical studies for two-dimensional quantum magnets have demonstrated excellent agreement with unbiased approaches 
\cite{Reuther2010,Reuther2011,Reuther2011a,Reuther2011b,Reuther2011c,Reuther2012,Reuther2014,Reuther2014a,Suttner2014,Rousochatzakis2015,Iqbal2015,Iqbal2016b}. 
The conceptual challenge to explain the deeper merits of the pf-FRG approach has recently been picked up by formulating a generalization of the pf-FRG to quantum magnets with arbitrary spin length $S$ \cite{Baez2017}. This spin-$S$ generalization becomes exact in the $S\to\infty$ limit where it precisely equals the well-known Luttinger-Tisza approach \cite{Luttinger1951,Luttinger1946}. This observation readily provides an explanation as to why the pf-FRG has been quite successful in detecting non-trivial magnetic ordering tendencies in systems with many competing interactions. 

In this manuscript, we introduce a generalization of the pf-FRG approach to SU($N$) quantum magnets. We show that the approach becomes exact also in the $N\to\infty$ limit. This not only complements the previous spin-$S$ generalization,
but it also provides an understanding why the pf-FRG approach has been quite successful in also identifying spin liquid regimes that do not show any magnetic ordering tendency. Such a suppression of magnetic ordering can be systematically studied in SU($N$) quantum magnets where quantum fluctuations are augmented by enlarging the spin symmetry group.
This 
leads to a simplification as these SU($N$) systems become exactly solvable in terms of a mean-field analysis in the limit $N\to\infty$. Generalized spin models of this kind have been used in the past as a starting point for a systematic expansion in $1/N$-corrections in an attempt to make statements about the physically prevalent (but far-away) $N=2$ case \cite{Arovas1988}. 
As we will discuss in the following, our SU($N$) pf-FRG generalization provides a systematic connection between these two limiting cases of $N=2$ and $N\to\infty$. We demonstrate this by a case study of the square-lattice SU($N$) Heisenberg antiferromagnet, which is well known to exhibit a N\'eel ordered ground state for SU(2) spins \cite{Manousakis1991},
while it harbors a staggered flux spin liquid in the $N\to\infty$ limit \cite{Arovas1988}. 
Our numerical implementation of the generalized SU($N$) pf-FRG approach shows that moderately enlarging the spin symmetry group readily destroys the formation of magnetic long-range order and allows us to track the formation of the staggered flux spin liquid at intermediate values of $N$. 
In addition, we find that the system, for sufficiently large values of $N$, develops a novel instability  that indicates the 
 transition into a spin liquid ground state. 
 This direct observation of a spin liquid transition in the pf-FRG framework is the first time that a {\em positive} identification of a spin liquid has been achieved with this approach. Within the real-space pf-FRG perspective on SU($N$) spin models introduced in this manuscript, the spin liquid phase below the phase transition, however, remains inaccessible. In a companion manuscript \cite{Roscher2017} we develop a momentum-space pf-FRG approach to SU($N$) spin systems that allows us to explicitly enter the spin liquid phase.
 In combination, these SU($N$) generalizations of the pf-FRG approach mark an important step towards closing the gap between the ability of existing numerical approaches for frustrated magnets and the conceptual understanding of spin liquids. 

The remainder of the manuscript is organized as follows. In Section \ref{sec:model} we introduce the SU($N$) Heisenberg model on the square lattice that we use as a case study to illustrate our generalized pf-FRG approach and to benchmark our results in the limiting cases of $N=2$ and $N \to \infty$ (where no $1/N$ corrections are included).
We provide a precise specification of the generalization of ordinary SU(2) quantum spins to SU($N$)-symmetric moments and briefly recapitulate the known mean-field results in the $N \to \infty$ limit. 
In Sec.~\ref{sec:SUN} we then turn to the generalized pf-FRG approach. We first derive the full set of pf-FRG flow equations in their SU($N$)-generalized form and discuss, on a formal level, the exact limiting case of large $N$. 
We numerically solve the flow equations for arbitrary $N$ in Sec. \ref{sec:finiteN} before we turn to an analytical solution of the $N\to\infty$ equations in Sec.~\ref{sec:largeN}. We review the breakdown of the solution at the spin-liquid phase transition and reconcile it with the established notion of a pf-FRG breakdown at an ordinary magnetic ordering transition.
We conclude our findings in Sec. \ref{sec:conclusions}.


\section{The SU(\textit{N}) square-lattice Heisenberg model}
\label{sec:model} 
While the generalized pf-FRG formalism developed in this manuscript can readily be applied to arbitrary SU($N$)
spin models in both two and three-dimensional lattice geometries, we consider for the sake of simplicity the SU($N$) Heisenberg model on the square lattice as a case study. 
Its Hamiltonian reads
\begin{equation}
\label{eq:hamiltonian}
H=\frac{1}{N}\sum\limits_{\langle i,j \rangle} J_{ij}~ {\bf S}_i \cdot {\bf S}_j \,,
\end{equation}
where the sum runs over nearest neighbors on the square lattice, and the exchange coupling is set to
be antiferromagnetic for all bonds $J_{ij}\equiv J \equiv 1$.
The  spin operators have been promoted to representations of the SU($N$) group. 
Note that there is no unique SU($N$) generalization of the spin algebra and, in fact, various representations of the generalization exist, both fermionic and bosonic, that may even describe different physical ground states in the large-$N$ limit \cite{Arovas1988}. Yet all schemes make a well-defined connection to the conventional SU(2) model. 
For the purpose of our work it is most convenient to choose a fermionic representation that expresses each spin in terms of $N$ different flavors of fermions, 
\begin{equation}
\label{eq:pseudofermions}
S_i^\mu=\sum\limits_{\alpha,\beta=1}^N f^\dagger_{i\alpha}T^\mu_{\alpha\beta}f^\nodagger_{i\beta} \,,
\end{equation}
where the $T^\mu$ ($\mu=1,\dots ,N^2-1$) are generators of the SU($N$) group. This is a valid spin representation if the following holonomic constraint is fulfilled, which corresponds to half-filling of fermions on each individual lattice site, 
\begin{equation}
\label{eq:fillingconstraint}
\sum_{\alpha=1}^N f^\dagger_{i\alpha} f^\nodagger_{i\alpha}=N/2 \text{~~for all~}i \,.
\end{equation}
This readily implies that the mapping is only well-defined for $N$ even. For $N=2$ this reduces to the conventional SU(2) representation in terms of Abrikosov fermions and Pauli matrices. 
Note that despite being a faithful representation of the spin operators the mapping from spins to fermionic operators introduces an artificial U(1) gauge symmetry \footnote{Note that the artificial gauge symmetry technically is SU(2), see Ref. \onlinecite{Affleck1988}. To clearly set this apart from the global SU(2) symmetry of the spin Hamiltonian we only address the implied U(1) character of the gauge symmetry.} that plays an important role in our analysis. 

\begin{figure}[t]
\centering
\includegraphics[width=0.95\linewidth]{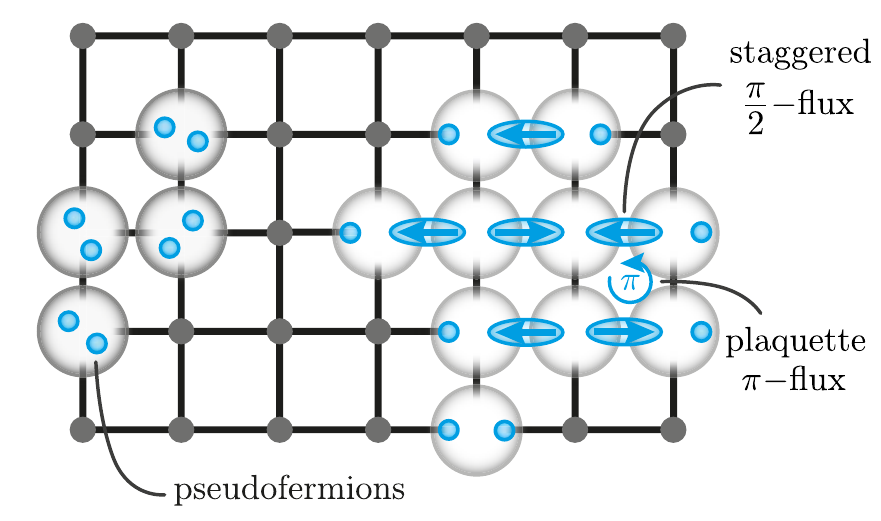}
\caption{{\bf SU(\textit{N}) spin model} for the staggered flux spin liquid.
The original spins are decomposed into pseudofermions and recombine into a finite, non-local order parameter on the bonds that generates a $\pi$-flux on every plaquette. }
\label{fig:model}
\end{figure}

A mean-field analysis provides us with important insight about the ground state of the system in the large-$N$ limit \cite{Arovas1988, Roscher2017}. 
Most importantly, a divergence of the magnetic susceptibility 
\begin{equation}
\label{eq:susceptibility}
\chi_{ij}=\int_0^\infty d\tau~\langle S_i^\mu(\tau) S_j^\mu(0) \rangle
\end{equation}
(no summation over $\mu$) is not observed which hints at the absence of magnetic long-range order. 
Instead one finds that the uniform susceptibility 
\begin{equation}
\chi(T)=\frac{1}{\mathcal{N}_L}\sum\limits_{ij} \chi_{ij}(T) \,,
\end{equation}
where $\mathcal{N}_L$ is the number of lattice sites, grows as $\chi(T)=1/(4T)$ above a critical temperature $T_c=J/4$. 
Below $T_c$ the susceptibility decreases again and eventually drops to zero. 
At the critical temperature one finds that the spins fractionalize and give rise to an emergent non-local field $Q_{ij}\sim f^\dagger_{i\alpha}f^\nodagger_{j\alpha}$ that is defined on the bonds of the original lattice. 
This order parameter develops a non-trivial spatial structure 
\begin{equation}
Q_{ij}=Qe^{-i\theta} \text{~~with~} \theta=
\begin{cases}
\pm\frac{\pi}{2} & \text{for~~} \mathbf{r}_j-\mathbf{r}_i\sim \mathbf{e}_x \\
0 & \text{for~~} \mathbf{r}_j-\mathbf{r}_i\sim \mathbf{e}_y 
\end{cases} \,,
\end{equation}
which becomes finite upon the transition into the staggered flux phase \cite{Affleck1988-2}, see Fig.~\ref{fig:model}. 
Note that the staggered flux spin liquid spontaneously breaks the U(1) symmetry\footnote{Here, we follow previous literature in employing the term ``symmetry breaking'' in the context of local gauge symmetries and use it analogous as, e.g., for mean field superconductors: There is no spontaneous symmetry breaking witnessed by correlators that are not gauge invariant, as a consequence of Elitzur's theorem. However, there are observable consequences due to the onset of ordering witnessed by gauge-invariant quantities, such as the expression of a gap in a superconducting phase.} and should be distinguished from the $\pi$-flux phase which comprises \emph{all} gauge equivalent configurations. 
With the non-local pairing of fermions the back transformation from pseudofermions to spins becomes somewhat ambiguous. In fact, the transformation is formally impaired by the release of the half-filling constraint \eqref{eq:fillingconstraint}. Even if the constraint was enforced by a Lagrange multiplier on the Hamiltonian level the corresponding prefactor would become zero in the mean-field solution \cite{Arovas1988}.


\begin{figure*}[t]
\centering
\includegraphics[width=0.8\linewidth]{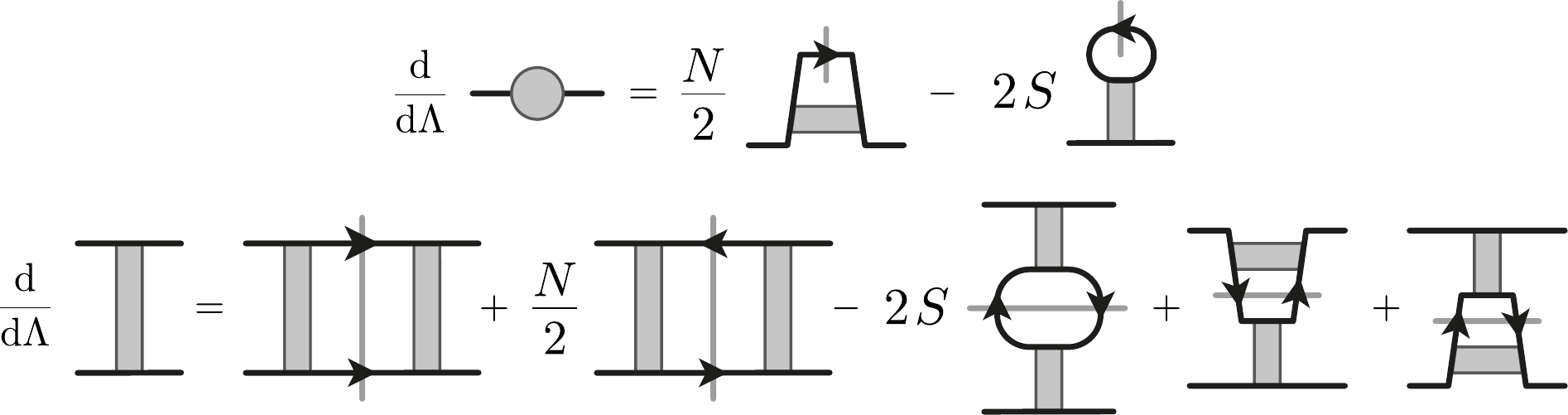}
\caption{{\bf Generalized flow equations} for SU($N$)-symmetric spin systems of spin length $S$. The prefactors indicated here are only schematic and to leading order in $S$ and $N$. The exact prefactors are given in Fig.~\ref{fig:flowequationsFull} of the appendix. Slashed propagator lines should be understood as the single-scale propagator. A pair of slashed propagators in the flow equation for the two-particle vertex should be understood as a pair of single-scale and full propagators where the two permutations of the propagators are summed over.
The Fock diagram and the particle-hole ladder contribution, which both describe quantum fluctuations are enhanced as $N$ is increased. Similarly, as the classical limit is approached for increasing spin length $S$ the Hartree and RPA channels are boosted. }
\label{fig:flowequation}
\end{figure*}


\section{SU(\textit{N})-generalization of pf-FRG}
\label{sec:SUN}

For arbitrary $N$, the mean field approach no longer remains exact and in fact introduces an unwanted bias.
Any mean-field decoupling reflects a choice of coupling channel that, without any prior understanding of the system,  
might introduce a preference  towards either the appearance of a spin liquid or a magnetically ordered state.
For an unbiased analysis it is, of course, more desirable to treat all channels on equal footing. 
This is precisely what the pf-FRG scheme \cite{Reuther2010}  allows for by addressing the full quartic interaction vertices
and thereby incorporating all possible decoupling channels in the RG flow. 
In the following, we will demonstrate how to set up the pf-FRG calculations for SU($N$) spin models by explicitly 
deriving the flow equations of the single- and two-particle interaction vertices (in their real-space representation).
Subsequently, we will discuss the relevance of the Katanin truncation \cite{Katanin2004} to the generalized flow equations
and show that this truncation scheme becomes exact in the limit $N \to \infty$. 
We close this Section with a short summary of our numerical implementation of the generalized pf-FRG scheme.


\subsection{Generalized flow equations}
\label{sec:FlowEquations}

Our starting point for the derivation of the generalized flow equations is the SU($N$) Hamiltonian \eqref{eq:hamiltonian}. Employing the fermionic representation of SU($N$) moments \eqref{eq:pseudofermions} we obtain the pseudofermionic Hamiltonian
\begin{equation}
\label{eq:hamiltonian2}
H=\frac{1}{N}\sum_{\langle i,j\rangle} J_{ij} T^\mu_{\alpha\beta}T^\mu_{\gamma\delta}  f^\dagger_{i\alpha}f^\nodagger_{i\beta}f^\dagger_{j\gamma}f^\nodagger_{j\delta} \,,
\end{equation} 
which is structurally equivalent to the pseudofermion Hamiltonian in the established pf-FRG scheme for SU(2)-symmetric spins \cite{Reuther2010}, i.e. it contains only quartic interactions and no kinetic terms. 
Therefore, in close analogy to the SU(2) pf-FRG scheme, we parametrize the self-energy as
\begin{equation}
\label{eq:selfenergyParm}
\Sigma^\Lambda(1';1)=\Sigma^\Lambda(\omega_1)\delta(\omega_{1'}-\omega_1)\delta_{i_{1'}i_1}\delta_{\alpha_{1'}\alpha_1} \,,
\end{equation}
and the two-particle vertex as 
\begin{equation}
\label{eq:effectActionParm}
\begin{split}
\Gamma^\Lambda(1',2';1,2)=\Big( \big[
\Gamma^\Lambda_{s,i_1i_2}(\omega_{1'},\omega_{2'};\omega_1,\omega_2) T^\mu_{\alpha_{1'}\alpha_1} T^\mu_{\alpha_{2'}\alpha_2} \\
+ \Gamma^\Lambda_{d,i_1i_2}(\omega_{1'},\omega_{2'};\omega_1,\omega_2) \delta_{\alpha_{1'}\alpha_1} \delta_{\alpha_{2'}\alpha_2}
\big] \delta_{i_1i_1'}\delta_{i_2i_2'} \\
- (1' \leftrightarrow 2')\Big)\times \delta(\omega_{1'}+\omega_{2'}-\omega_1-\omega_2) \,,
\end{split}
\end{equation}
where $\Lambda$ is the frequency cutoff scale and the numbers $n$ in the vertex arguments represent composite indices of lattice site $i_n$, spin index $\alpha_n$ (that runs from 1 to $N$), and Matsubara frequency $\omega_n$.  
This parametrization of the effective action is a complete basis for all SU($N$)-symmetry allowed contributions -- comprising a spin term 
\[\Gamma_s\propto T^\mu_{\alpha_1\alpha_1'}T^\mu_{\alpha_2\alpha_2'}\] 
and a density term \[\Gamma_d\propto \delta_{\alpha_1\alpha_1'}\delta_{\alpha_2\alpha_2'}\,.\]
The initial values for the vertices at infinite cutoff $\Lambda\to\infty$ are given by the bare interactions and may be obtained from comparison with the Hamiltonian \eqref{eq:hamiltonian2}: 
\begin{equation}
\begin{split}
&\Sigma^\infty(\omega)=0 \,,\\
&\Gamma^\infty_{s,i_1i_2}(\omega_{1'},\omega_{2'};\omega_1,\omega_2)=J_{i_1i_2}/N \,,\\
&\Gamma^\infty_{d,i_1i_2}(\omega_{1'},\omega_{2'};\omega_1,\omega_2)=0 \,.
\end{split}
\end{equation}
Note that even though the density contribution is zero initially it may become finite throughout the flow and therefore must be included in the calculation. 
For the pf-FRG scheme to be complete we have to specify an RG-cutoff function which is conveniently chosen as a sharp cutoff in frequency space \cite{Reuther2010}. 
With this cutoff function the full propagator and the single-scale propagator, respectively, become 
\begin{align}
\label{eq:propagator}
G^\Lambda(\omega)&=\frac{\Theta(|\omega|-\Lambda)}{i\omega-\Sigma^\Lambda(\omega)} \,,\\
\label{eq:singlescalepropagator}
S^\Lambda(\omega)&=\frac{\delta(|\omega|-\Lambda)}{i\omega-\Sigma^\Lambda(\omega)} \,.
\end{align}
The full set of flow equations can then be derived by inserting the parametrization of the effective action (\ref{eq:selfenergyParm},\ref{eq:effectActionParm}) together with the propagators \eqref{eq:propagator} and \eqref{eq:singlescalepropagator} into the general form of the fermionic flow equations \cite{Metzner2012} 
\begin{align}
\frac{d}{d\Lambda}\Sigma^\Lambda(1';1)&=-\frac{1}{2\pi}\sum\limits_2 \Gamma^\Lambda(1',2;1,2)S^\Lambda(\omega_2) \,,\\
\frac{d}{d\Lambda}\Gamma^\Lambda(1',2';1,2)&=\frac{1}{2\pi}\sum\limits_{3,4} 
[\Gamma^\Lambda(1',2';3,4)\Gamma^\Lambda(3,4;1,2) \nonumber\\
&-\Gamma^\Lambda(1',4;1,3)\Gamma^\Lambda(3,2';4,2) - (3\leftrightarrow 4) \nonumber\\
&+\Gamma^\Lambda(2',4;1,3)\Gamma^\Lambda(3,1';4,2) + (3\leftrightarrow 4)] \nonumber\\
&\times G^\Lambda(\omega_3)S^\Lambda(\omega_4) \,.
\end{align}
The general structure \footnote{The spin structure of the flow equations can be solved exactly giving two separate flow equations for the spin contribution $\Gamma^\Lambda_s$ and the density contribution $\Gamma^\Lambda_d$ of the two-particle vertex. 
}
of these generalized flow equations is depicted in Fig.~\ref{fig:flowequation}. 
As $N$ is increased, the Fock channel in the flow of the single-particle vertex and the particle-hole channel in the flow of the two-particle vertex become dominant. This should be contrasted to what has been observed in the large-$S$ generalization of the flow equations \cite{Baez2017}, where the Hartree and RPA channels become dominant. 
The full set of SU($N$)-symmetric flow equations for arbitrary $N$ is given in the appendix.

Since the flow equations provide a full description of all single-particle and two-particle interaction vertices they can readily be used to compute observables. The magnetic susceptibility \eqref{eq:susceptibility}, in particular, can be diagrammatically expanded as 
\begin{equation}
\vcenter{\hbox{\includegraphics[scale=1]{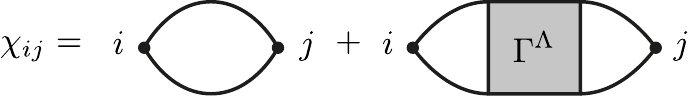}}} \,.
\end{equation}
The so-calculated susceptibility is often used as primary witness for the presence or absence of magnetic order via the presence or absence of a flow breakdown as we will discuss in further detail in Section \ref{sec:GeneralConsiderations}. 
To reveal the incipient magnetic order in case of a flow breakdown, one typically calculates real-space correlations and their momentum-space structure factor in the vicinity of the flow breakdown.


\subsection{Katanin truncation}
\label{sec:Katanin}

It is important to note that the most general form of the fermionic FRG flow equations is an infinite hierarchy of integro-differential equations where the flow of the $n$-particle vertex may depend on all vertices up to order $n+1$. In order to obtain a closed set of equations, the dependence of the flow equation for the 2-particle vertex on the 3-particle vertex has been truncated in the above discussion. 
However, it has been shown phenomenologically that, at least for SU(2) spin models, this simple truncation is not sufficient to predict the correct physical behavior of many spin models \cite{Reuther2010}. On this rudimentary level of truncation, the flow of the magnetic susceptibility would always diverge at some finite cutoff scale $\Lambda_c$, which can be related to a spontaneous breaking of spin-rotational symmetry that is explicitly incorporated in the vertex parametrization \eqref{eq:effectActionParm}. Once the symmetry is spontaneously broken, the vertex parametrization and hence also the flow equations can no longer describe the correct physical behavior -- which is typically observed in the form of a divergence or a kink in the RG flow of the magnetic susceptibility. For an arbitrary spin system, its magnetic ordering tendencies are thereby highly overestimated, while spin-liquid phases cannot be observed at all. 
The crucial step to overcome this limitation is to improve the truncation by implementing the so-called Katanin scheme \cite{Katanin2004}, which stipulates the replacement of the single-scale propagator with the derivative of the full propagator, 
\begin{equation}
\label{eq:katanin}
S^\Lambda(\omega)\to S^\Lambda_\mathrm{kat}(\omega)=-\frac{d}{d\Lambda}G^\Lambda(\omega) \,.
\end{equation}
Making this replacement was phenomenologically shown  to be a vital ingredient in order to correctly predict spin liquid phases \cite{Reuther2010} in various SU(2) spin models. 

It is thus obvious to also implement the Katanin truncation for the generalized SU($N$) flow equations. Importantly, we will explicitly show that the Katanin truncation is in fact a necessary ingredient to exactly reproduce the mean-field gap equation in the large-$N$ limit~\cite{Salmhofer2004}. But before we proceed to such a rigorous algebraic analysis, we want to first lay out an intuitive diagrammatic understanding of the truncation scheme for the large-$N$ calculations.
To simplify the diagrammatics we rewrite the pseudofermionic Hamiltonian \eqref{eq:hamiltonian2} 
\begin{equation}
H=\frac{1}{2N}\sum_{\langle i,j\rangle} J_{ij} f^\dagger_{i\alpha}f^\nodagger_{i\beta}f^\dagger_{j\beta}f^\nodagger_{j\alpha} \,,
\end{equation}
which to leading order in $1/N$ is equivalent (up to a factor of 2 that can be absorbed into the definition of the coupling constant) to the previous formulation that explicitly revealed the spin exchange $\propto T^\mu_{\alpha_1\alpha_1'}T^\mu_{\alpha_2\alpha_2'}$. 
Diagrammatically, the bare interactions can now be represented as 
\begin{equation}
\vcenter{\hbox{\includegraphics[scale=0.8]{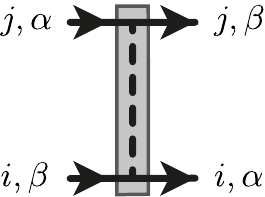}}}\sim \frac{1}{N} \,.
\end{equation}
The flow equation in its truncated form (that we discussed in the previous section, where the three-particle vertex has been set to zero) contains, to leading order in $1/N$, only a single contribution from the particle-hole channel (c.f. Fig.~\ref{fig:flowequationsFull} in the appendix), 
\begin{equation}
\label{eq:kat_flowequation}
\vcenter{\hbox{\includegraphics[scale=0.8]{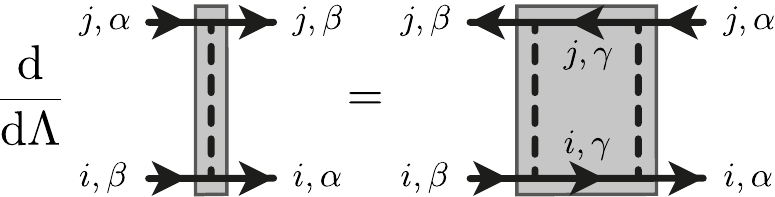}}} \,.
\end{equation}
The full flow equation for the two-particle vertex, however, also contains a contribution from the three-particle vertex, where two external legs are contracted. This contribution may indeed become finite to leading order in $N$, e.g. generated by the following SU($N$)-symmetric vertex $\sim~f^\dagger_{i\alpha} f^\nodagger_{i\beta} f^\dagger_{j\beta} f^\nodagger_{j\alpha} f^\dagger_{i\gamma} f^\nodagger_{i\gamma}$ where two external legs are contracted,
\begin{equation}
\label{eq:kat_threeparticle}
\vcenter{\hbox{\includegraphics[scale=0.8]{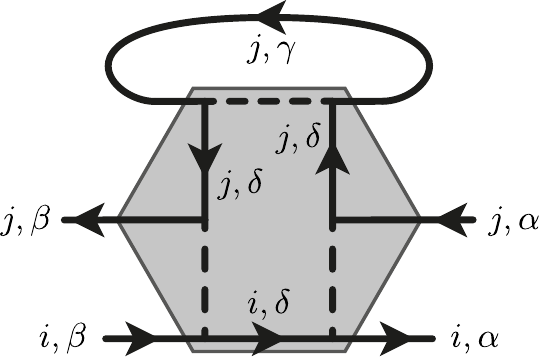}}} ~\sim~ \frac{N^2}{N^3} \,.
\end{equation}
Although such terms are not included in a straight-forward truncation of the three-particle vertex, they are captured by the Katanin scheme:  
Diagram \eqref{eq:kat_threeparticle} is just a particle-hole ladder diagram where one internal propagator has been replaced by a Fock diagram (c.f. the first diagram in the single-particle flow equation, Fig. \ref{fig:flowequation}). This is exactly what the Katanin scheme \eqref{eq:katanin} prescribes -- the replacement of the single-scale propagator with the derivative of the full propagator, which in the large-$N$ limit includes a Fock diagram contribution. 

Higher orders, however, are not captured by the Katanin scheme. Contributions from the four-particle vertex (with four out of eight external legs contracted) or higher orders can also be constructed to leading order in $N$ that cannot be reduced to the particle-hole channel with a single propagator substitution. 
It is therefore not immediately clear that the Katanin truncation suffices to describe the exact large-$N$ behavior. This is easier to see by formulating an implicit solution to the Katanin truncated flow equations. 

To this end, let us consider the flow equations in their Katanin-truncated form. The numbers now represent composite indices of Matsubara frequency and lattice site while the spin structure has been computed explicitly, and we only keep terms to leading order in $N$,
\begin{align}
\label{eq:kat_flowselfenergy}
\frac{d}{d\Lambda}\Sigma^\Lambda(1';1)&=\frac{1}{2\pi}\sum\limits_{22'} \Gamma^\Lambda(1',2';1,2)S^\Lambda(2;2') \,,\\
\frac{d}{d\Lambda}\Gamma^\Lambda(1',2';1,2)&=-\frac{1}{2\pi}\sum\limits_{33',44'} \frac{d}{d\Lambda} \left[ G^\Lambda(3;3')G^\Lambda(4;4') \right] \nonumber\\
&\times \Gamma^\Lambda(1',4';1,3)\Gamma^\Lambda(3',2';4,2) \,.
\end{align}
Notice how the Katanin replacement allows us to combine two terms in the two-particle flow equation into a single derivative of a product of two propagators. 
For the conventional truncation such a concise notation is not possible, and both terms appear explicitly, 
\begin{align}
\frac{d}{d\Lambda}&\Gamma^\Lambda(1',2';1,2)= \nonumber\\
 &-\frac{1}{2\pi}\sum\limits_{33',44'} \big[ S^\Lambda(3;3')G^\Lambda(4;4') + G^\Lambda(3;3')S^\Lambda(4;4') \big] \nonumber\\
 &\times \Gamma^\Lambda(1',4';1,3)\Gamma^\Lambda(3',2';4,2) \,.
\end{align}
This concatenation of terms is one of the biggest perks of the Katanin scheme. It allows us to formulate an implicit solution of the flow equations \cite{Salmhofer2004}. The solution is constructed as 
\begin{align}
\label{eq:kat_solution}
\Gamma^\Lambda(1',2';1,2) =~ &\Gamma^\infty(1',2';1,2) \nonumber\\
&- \frac{1}{2\pi} \sum\limits_{33',44'} G^\Lambda(3;3')G^\Lambda(4;4')  \nonumber\\
& \times \Gamma^\infty(1',4';1,3)\Gamma^\Lambda(3',2';4,2)
\end{align}
and describes a resummation of spin loops that ensures relevance to leading order in $N$. Inserting the solution \eqref{eq:kat_solution} into the flow equation for the self energy \eqref{eq:kat_flowselfenergy}, one obtains the relation
\begin{equation}
\frac{d}{d\Lambda}\Sigma^\Lambda(1';1)=-\frac{1}{2\pi}\sum\limits_{22'} \Gamma^\infty(1',2';1,2) \frac{d}{d\Lambda}G^\Lambda(2;2')
\end{equation}
which due to vanishing initial value of the propagator is straight-forward to integrate. 
We obtain a self-consistent expression for the non-local self-energy
\begin{equation}
\Sigma^\Lambda(1';1)=-\frac{1}{2\pi}\sum\limits_{22'} \Gamma^\infty(1',2';1,2)G^\Lambda(2;2')
\end{equation}
that reproduces the self-consistent gap equation in the mean-field formalism. 


\subsection{Numerical solution of the RG equations}
\label{sec:NumericalSolution}

We round off this Section by providing some details on the numerical solution of the pf-FRG flow equations for arbitrary SU($N$) spin models. Formally, we consider the flow equations in their zero temperature limit, where Matsubara frequencies are continuous. Hence, for the numerical solution of the flow equations one has to artificially discretize the vertices' frequency dependence. To do so, we typically introduce a logarithmic frequency mesh with some $\mathcal{N}_\omega=144$ discrete frequencies. 
Furthermore, we note that the flow equations include summations over the entire real-space lattice (Hartree and RPA term, see Fig.~\ref{fig:flowequation}). To treat such terms numerically, we consider interactions only up to a certain bond-distance and truncate interactions beyond this. Note that this scheme does not introduce an artificial boundary to the system and should rather be understood as a finite-cluster expansion operating directly in the thermodynamic limit. 
Convergence is usually reached already for distances of $L=10$ bonds in any direction (which for the square-lattice model at hand  corresponds to a cluster of $\mathcal{N}_L=221$ lattice sites), even for phases of magnetic long-range order. 
This leaves us -- after employing all lattice symmetries -- with a total of 13,623,624 coupled differential equations to solve.


\section{Results}
\label{sec:results}

\subsection{General considerations}
\label{sec:GeneralConsiderations}

Before we turn to explicit numerical results for the solution of the generalized flow equations, let us start with some
general considerations. In Section \ref{sec:FlowEquations}  we already mentioned that the principal signature of
a magnetic ordering transition, where the SU($N$) spin symmetry is spontaneously broken, is a breakdown of the 
smooth flow of the magnetic susceptibility \cite{Reuther2010}. 
In the following, we recapitulate \cite{JensHabil} that not only a magnetic ordering transition but, more generally, any second order phase transition necessarily results in a breakdown of the flow -- given that it is driven by an interaction that is quartic in fermions. The argument holds in particular for the transition into the staggered flux spin liquid that we expect to emerge in our model system \eqref{eq:hamiltonian} and which is accompanied by a spontaneous breaking of the U(1) symmetry. 
To this end let us consider the action of a generic four-fermion exchange term
\begin{equation}
\label{eq:hst_interaction}
S_\mathrm{int} = J\sum_{1234}f^\dagger_1 f^\nodagger_2f^\dagger_3 f^\nodagger_4 \,,
\end{equation}
where the indices are composite symbols for any relevant set of quantum numbers and $J$ is the (constant) interaction strength. Note that the generalized case $J = J_{1234}$ can be addressed by considering each realization of $J_{1234}$ separately.
Now suppose that the action $S = S_0 + S_\mathrm{int}$ with the non-interacting part $S_0$ is invariant under some symmetry group $G$, but the ground state of the system is known to spontaneously break this symmetry -- assume, for example, $G=\mathrm{U(1)}$, which is the symmetry spontaneously broken by the staggered flux spin liquid. 
Let us consider a bosonic order parameter $Q^{(\dagger)}_{12}$ whose onset is an indicator of the spontaneous symmetry-breaking. By means of a Hubbard-Stratonovich transformation it is then possible to reformulate the partition function $\mathcal{Z}$ such that the order parameter field is exposed explicitly, i.e.
\begin{equation}
\mathcal{Z} = \int\mathcal{D}[f]~e^{-S} = \mathcal{N}\int\mathcal{D}[f,Q]~e^{-S - m \sum_{1234}Q^\dagger_{12} Q^\nodagger_{34}} \,.
\end{equation}
For concreteness, but without loss of generality, we shall assume that the ordering occurs in a density-like channel where a linear shift of the form
\begin{equation}
Q^{(\dagger)}_{12}\,\rightarrow\, Q^{(\dagger)}_{12} + \frac{g}{m}f^\dagger_1 f^\nodagger_2
\end{equation}
may be employed. If the defining relation
\begin{equation}
\label{eq:hst_definingrelation}
m = \frac{g^2}{J}
\end{equation}
is fulfilled, the quartic fermion interaction $S_\mathrm{int}$ is canceled exactly and the partition function takes on the form
\begin{equation}
\label{eq:hst_transformedaction}
\mathcal{Z} = \int\mathcal{D}[f,Q]~e^{-S_0 - \sum_{1234}\left[ m Q^\dagger_{12} Q^\nodagger_{34} + g f^\dagger_1 f^\nodagger_2 Q^{\phantom\dagger}_{34} + g Q^\dagger_{12} f^\dagger_3 f^{\phantom\dagger}_4 \right]} \,.
\end{equation}

Let us now discuss the implications for our RG analysis and treat the coefficient $m\equiv m^\Lambda$ of the ($G$-invariant) order parameter density $\rho = Q^\dagger Q$ as a running coupling that depends on the RG cutoff $\Lambda$. Using Eq. \eqref{eq:hst_transformedaction} as the initial condition for the flow of the effective average action, higher powers of $\rho$ are generated throughout the flow and eventually give rise to an order parameter potential $U[\rho]$ in the Landau sense. 
In the Landau picture, the spontaneous symmetry breaking by means of a second order phase transition is tied to a sign change of the linear term in the potential $U[\rho]$ -- which is parametrized by the running coefficient $m^\Lambda$. 
Alternatively, the Hubbard-Stratonovich transformation may in principle be performed on the original fermionic formulation with the running coupling $J^\Lambda$ at any value for the cutoff $\Lambda$. Therefore relation~\eqref{eq:hst_definingrelation} generally holds throughout the entire RG flow (up to artifacts arising from the truncation of the flow equations). Hence, the sign change (and thus a zero-crossing) of $m^\Lambda$ in the bosonic formulation is equivalent to a divergence of the coupling $J^\Lambda$ in the fermionic model.

Thus, the spontaneous breaking of any symmetry group $G$ goes hand in hand with a divergence of the fermionic coupling $J$ as long as the order parameter field can be represented by some fermion bilinear. 
In our example, where $G=\mathrm{U(1)}$, and the symmetry-breaking order parameter is $Q_{ij}\sim f^\dagger_{i\alpha}f^\nodagger_{j\alpha}$, we therefore expect to observe a breakdown of the flow upon transition into the staggered flux spin liquid phase. 
Note that the bosonized description \eqref{eq:hst_transformedaction} requires explicit knowledge about the structure of the order parameter. The purely fermionic formulation, however, is unbiased as the breakdown is bound to occur on the level of the effective average action without any additional transformation.

Having formulated the expectation of a vertex divergence at the flow-induced phase transition, we are almost set do discuss the solution of the flow equations. But before we proceed a remark on our strategy for obtaining {\em finite-temperature} observables is in order.
That is because formally we always solve the flow equations at zero temperature, and we refrain from introducing a finite temperature in the sense of a Matsubara frequency discretization. 
Solving the zero-temperature flow equations we obtain the full dependence of the susceptibility $\chi^\Lambda$ (or any other observable) on the cutoff parameter $\Lambda$. Per construction of the flow equations the zero-cutoff limit describes the true physical solution of the system and in our case corresponds to the ground state at zero temperature. 
However, it has been shown heuristically that -- at least in three spatial dimensions, where a finite-temperature transition exists -- the cutoff dependence of the formal zero-temperature solution can be related to the full temperature dependence of the observable by a simple rescaling \cite{Iqbal2016},
\begin{equation}
\label{eq:rescalingrelation}
\chi(T)=\chi^\Lambda |_{\Lambda=2T/\pi} \,.
\end{equation} 
In the remainder of this section we shall see that the relation also holds for our two-dimensional model system above the phase transition into the staggered flux spin liquid. In the symmetry-broken regime the simple relation \eqref{eq:rescalingrelation} no longer holds (see our companion manuscript of Ref.~\onlinecite{Roscher2017} for a more detailed discussion).

\subsection{Finite-\textit{N} results}
\label{sec:finiteN}

\begin{figure}
\centering
\includegraphics[width=\linewidth]{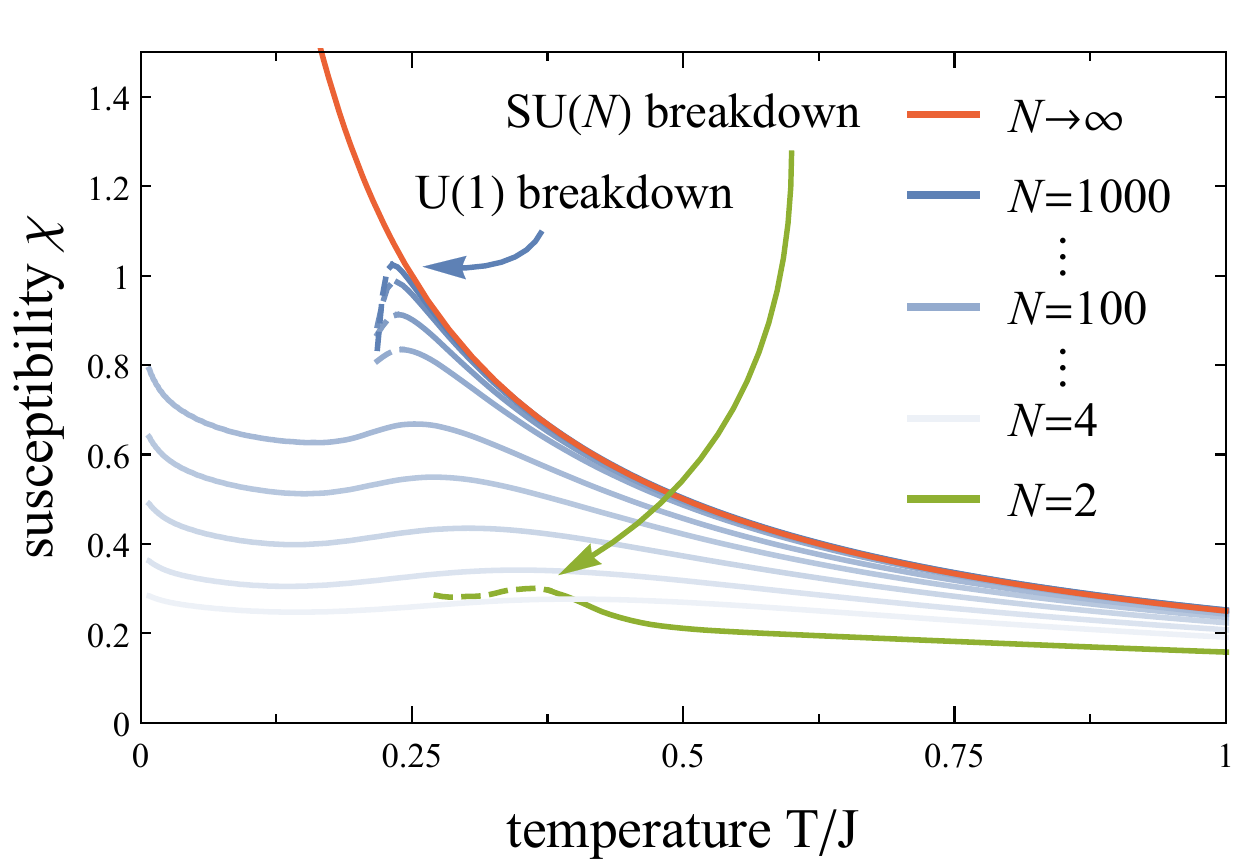}
\caption{{\bf Susceptibility for various \textit{N}} calculated by pf-FRG \emph{with} Katanin truncation. For finite $N$ the flow equations are solved numerically. In the limit $N\to\infty$ the flow equation is solved analytically.}
\label{fig:finiteNKat}
\end{figure}

\begin{figure}[b]
\centering
\includegraphics[width=\linewidth]{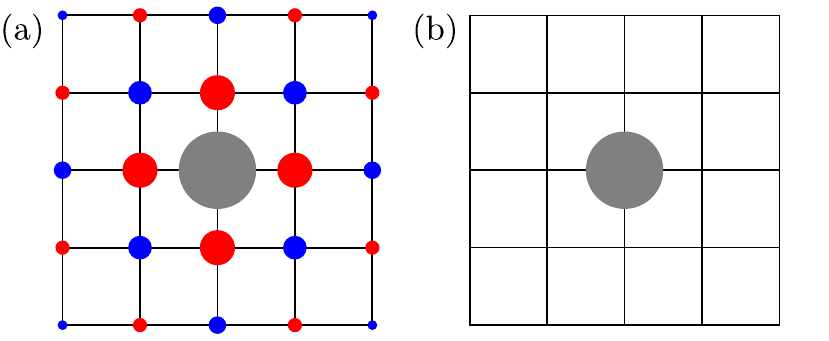}
\caption{{\bf Real space correlations} for (a) $N=2$ and (b) $N=1000$ plotted just above the respective critical temperature. Correlations are given with respect to a single reference site (grey). Ferromagnetic correlations are represented by blue circles, AFM correlations by red ones. The size of the circle represents the correlation's magnitude relative to the reference site. Correlations for $N=2$ clearly reveal N\'eel order while for $N=1000$ no correlations are visible.}
\label{fig:correlations}
\end{figure}

We now proceed to an analysis of the square-lattice SU($N$) Heisenberg model by a straight-forward numerical solution of the generalized flow equations for arbitrary $N$. By first considering the limiting case of $N=2$, we can convince ourselves that we can indeed reproduce the previous results for the ordinary SU(2) Heisenberg antiferromagnet, which is known to exhibit N\'eel order at zero temperature. 
Indeed we observe a breakdown in the flow of the uniform susceptibility in our pf-FRG calculations around $T\approx0.37$ (see Fig.~\ref{fig:finiteNKat}) that is attributed to a spontaneous breaking of the SU(2) spin symmetry upon magnetic ordering. 
Yet the specific value for a transition temperature should not be taken literally here; Despite the flow equations being formulated at zero temperature, in two spatial dimensions one should be careful about the rescaling from frequency cutoff to temperature since a finite-temperature transition arising from the spontaneous breaking of a continuous symmetry is formally excluded by the Mermin-Wagner theorem \cite{MerminWagner}.
As the full spatial dependence of the two-particle vertex is calculated in the pf-FRG scheme, we can directly access the real-space resolved correlations. Although we cannot access the symmetry-broken regime where true long-range N\'eel order prevails, we can study the susceptibility just above the critical temperature where the incipient magnetic order is already visible in the correlations. 
For the SU(2) model the antiferromagnetic correlations observed right above the transition clearly depict N\'eel order, as shown in the left panel of Fig.~\ref{fig:correlations}. 

Already a slight enlargement of the symmetry group to SU(4) is found to cause a qualitative change. Quantum fluctuations in the extended symmetry group appear sufficient to destroy magnetic order even at zero temperature. Consequently, the flow breakdown disappears and the susceptibility runs smoothly down to zero temperature without any sign of magnetic ordering,
see Fig.~\ref{fig:finiteNKat}. 
This intermediate regime persists until the spin symmetry is enlarged significantly to $N\approx 100$, where we identify a third qualitatively different regime. Notably, above $N\approx100$ a new flow breakdown develops around $T_c \approx 1/4$ as shown in Fig.~\ref{fig:finiteNKat}. 
These results are in line with previous Quantum Monte-Carlo studies that agree on the breakdown of N\'eel order at $N=6$, while the $N=4$ case is more subtle \cite{Assaad2005, Cai2013}.
Until now a breakdown of the flow has always been related to the onset of magnetic ordering in the framework of pf-FRG. Yet the breakdown that we observe here is of different origin. This becomes apparent already by a quick check of the spin correlations that are practically zero within the entire lattice (see right panel of Fig.~\ref{fig:correlations}). 
In fact, the breakdown should be interpreted as a signature of the spontaneous breaking of the artificial U(1) gauge symmetry that is introduced with the fermionization process \eqref{eq:pseudofermions} of the original spin model. 
This is in line with the mean-field result for the $N \to \infty$ limit, which predicts a transition into the U(1)-broken staggered flux spin liquid phase at $T_c=1/4$ and whose high-temperature susceptibility we perfectly reproduce for large, but finite $N$. 
Our numerical pf-FRG analysis thus suggests that the staggered flux spin liquid and the associated finite-temperature phase transitions persist for finite, but large $N$. A more detailed analysis of the symmetry-broken phase with a modified momentum-space pf-FRG approach to SU($N$) spin models is developed in the companion manuscript~\cite{Roscher2017}, which allows us to explicitly enter the low-temperature phase and explicitly probe symmetry-breaking properties.

\begin{figure}
\centering
\includegraphics[width=\linewidth]{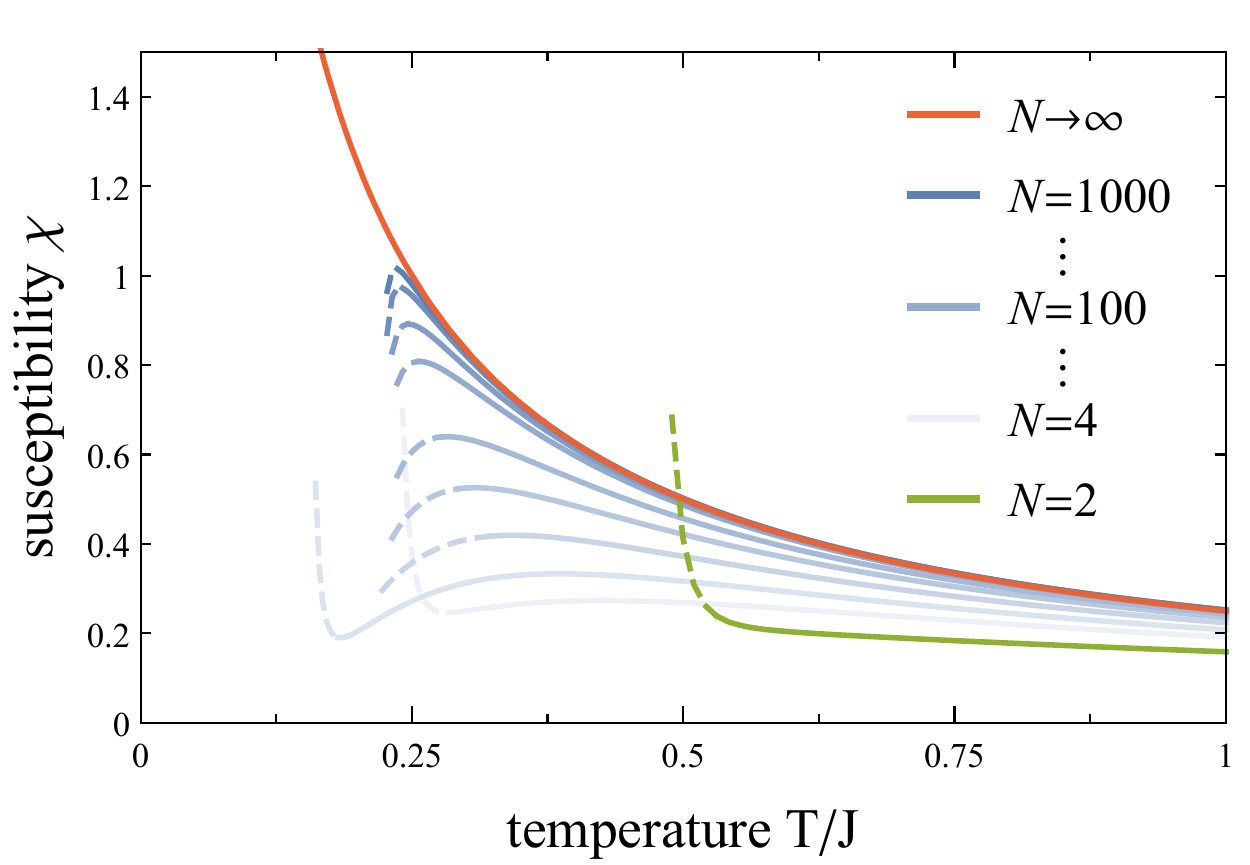}
\caption{{\bf Susceptibility for various \textit{N} \emph{without} Katanin truncation} calculated by pf-FRG. Finite-$N$ results are obtained numerically, the $N\to\infty$ limit is exact.}
\label{fig:finiteNNoKat}
\end{figure}

Knowing that for large $N$ we approach the exact result above the phase transition raises the question how important the Katanin scheme really is for our results. 
We have established in Sec. \ref{sec:Katanin} that the Katanin scheme is essential to faithfully reproduce the exact mean-field result in the large-$N$ limit. 
Comparing the finite-$N$ results that we obtain with (Fig.~\ref{fig:finiteNKat}) and without (Fig.~\ref{fig:finiteNNoKat}) the Katanin scheme we find that deviations only exist for $N\lesssim 100$, i.e. explicitly \emph{not} in the large-$N$ limit. 
For small $N=2$ the strong sensitivity of the result on the truncation scheme is known already \cite{Reuther2010}. Without the Katanin scheme the susceptibility would always diverge, making it impossible to identify spin liquid phases. 
In our results this behavior is particularly distinct for moderate $N=4,6,8,\ldots$ where the susceptibility runs smoothly down to zero temperature if the Katanin scheme is applied but the flow breaks down in the conventional truncation scheme. 
At large $N$ the Katanin scheme may still become relevant in the staggered flux spin liquid that we cannot access. 
Intuitively one would indeed expect that higher order diagrams, which are only captured within the Katanin scheme, become more important only in the symmetry-broken phase where the non-local order parameter $Q_{ij}$ becomes finite. 
Since such a term is not allowed in our initial parametrization of the self-energy \eqref{eq:selfenergyParm} we are oblivious of its existence. 
To explore the symmetry-broken phase we should therefore allow an explicit site-dependence in the propagator and replace $G(\omega)\to G_{ij}(\omega)$. 
It turns out, however, that this modification alone is not sufficient to capture the symmetry-broken phase. 
Throughout the flow the existence of a non-local propagator may also give rise to two-particle vertices with a more complicated spatial structure. 
To be consistent, one must therefore also allow for non-local two-particle terms $\Gamma_{s/d,i_1i_2}\to \Gamma_{s/d,i_1'i_2'i_1i_2}$ that carry a full site-dependence. 
With two additional lattice site indices per vertex, this generalized approach comes with a massive increase of computational complexity on the order of $\mathcal{N}_L^2$, and is not feasible numerically. 

A different pf-FRG approach which is condensed to a relatively small number of relevant flowing parameters is presented in our companion manuscript Ref.~\onlinecite{Roscher2017} and explicitly explores the symmetry-broken phase.
In the remainder of this manuscript we instead discuss the exact solution of the $N\to\infty$ limit to get a better understanding of the U(1)-symmetry breaking transition and its signatures in the pf-FRG formalism.


\subsection{Large-\textit{N} limit}
\label{sec:largeN}

For the strict $N\to\infty$ limit, which is fundamentally different from any finite $N$, special attention is in order. Most notably, it does not continuously connect to any large but finite value for $N$. 
The $N\to\infty$ limit no longer shows any sign of a flow breakdown in the susceptibility. Yet we have established already that this limit should be exact and that we expect the two-particle vertex to diverge. 
We can shed light onto this puzzle by solving the flow equations exactly, which is possible in the limit $N\to\infty$. 

Taking into account only the leading order terms in $1/N$ the flow equation for the spin-channel of the two-particle vertex only contains the particle-hole diagram which is local in the lattice site index, i.e.~only those vertices evolve that have a non-zero initial condition (c.f. flow equations in Fig. \ref{fig:flowequationsFull} of the appendix). 
The density channel remains strictly zero throughout the entire flow due to its initial conditions. 
The flow of the self energy  depends only on the Fock diagram, whose spatial structure is such that it can become non-zero only for a finite on-site two-particle interaction, which to leading order does not exist. 
Therefore, the only non-zero diagrams are the spin-channel interactions for nearest neighbor sites on the lattice where the initial value is already finite. 
From here on we will only consider those vertices and simplify notation by suppressing lattice site indices as well as using energy conservation to reduce the number of frequency arguments. Furthermore, we rescale all vertices by a factor of $N$ to eliminate the explicit $N$-dependence from the flow equations, 
\begin{equation}
\Gamma^\Lambda_{s,i_1i_2}(\omega_{1'},\omega_{2'};\omega_1,\omega_2) \to 
\frac{1}{N}\Gamma(\omega_{1'}+\omega_{2'},\omega_{1'}-\omega_1,\omega_{1'}-\omega_2) \,.
\end{equation}
With those simplifications we arrive at a compact expression for the flow equations of the spin channel of nearest neighbor interactions: 
\begin{equation}
\begin{split}
\frac{d}{d\Lambda}\Gamma^\Lambda(s,t,u)=-\frac{1}{4\pi}\int d\omega'  \\
\left[ \Gamma^\Lambda(\omega_2'-\omega',-\omega_1-\omega',u) \Gamma^\Lambda(\omega_2-\omega',\omega_1'+\omega',u) \right. \\
\left. \Gamma^\Lambda(\omega_1+\omega',-\omega_2'+\omega',u) \Gamma^\Lambda(\omega_1'+\omega',\omega_2-\omega_1',u) \right] \\
\times \frac{\delta(|\omega'|-\Lambda)}{\omega'} \frac{\Theta(|u+\omega'|-\Lambda)}{u+\omega'}
\end{split}
\end{equation}
with the initial condition 
\begin{equation}
\Gamma^{\Lambda\to\infty}(s,t,u)=J \,,
\end{equation}
and the transfer frequencies $s$, $t$, and $u$ are defined as
\begin{align}
s&=\omega_{1'}+\omega_{2'} \,, \nonumber\\
t&=\omega_{1'}-\omega_1 \,,\nonumber\\
u&=\omega_{1'}-\omega_2 \,.
\end{align}
Note that here the Katanin truncation is indeed equivalent to the conventional truncation, since all corrections by the Katanin scheme depend on the derivative of the self-energy, which is strictly zero. 
Since the initial condition for the vertices is constant in frequency space, the flow also remains constant in $s$ and $t$ direction. We therefore further simplify our notation and suppress those indices and  only explicitly state the non-trivial dependency on $u$. 
Solving the frequency integration over the delta- and theta-functions one obtains the simplified flow equation 
\begin{equation}
\frac{d}{d\Lambda}\Gamma^\Lambda(u)=-\frac{1}{2\pi} \frac{\left(\Gamma^\Lambda(u)\right)^2}{\Lambda(u+\Lambda)} \,,
\end{equation}
which is  solved in a straight-forward manner by 
\begin{equation}
\Gamma^\Lambda(u)=\frac{J}{1-\frac{J}{2\pi u}\log\left(1+\frac{u}{\Lambda}\right)} \,.
\end{equation}
Structurally, this result is exactly what we expected. The first vertex to diverge is the $u=0$ component at a critical cutoff scale $\Lambda_c=\frac{J}{2\pi}$. 
Applying the rescaling from frequency cutoff to temperature \eqref{eq:rescalingrelation}, the critical cutoff is equivalent to a critical temperature of $T_c=J/4$ and exactly matches the mean-field result. The divergence of a vertex tells us that we must not trust our solution below $T_c$.  

Yet the question remains why we cannot observe the vertex divergence in the magnetic susceptibility measurement. To understand this, we explicitly calculate the susceptibility which to leading order in $N$ only depends on the single-particle propagator 
\begin{equation}
\chi^\Lambda=-\frac{1}{4\pi}\int d\omega \left(G^\Lambda(\omega)\right)^2 \,.
\end{equation}
It is important here that the direct contribution of the two-particle vertex is suppressed in the large-$N$ limit, and since the self-energy remains strictly zero, the expression for the susceptibility becomes trivial, 
\begin{equation}
\chi^\Lambda=-\frac{1}{4\pi}\int d\omega \left(\frac{\Theta(|\omega|-\Lambda)}{w} \right)^2 \,.
\end{equation}
Solving the integral yields $\chi^\Lambda=({2\pi\Lambda})^{-1}$ which allegedly is valid for any $\Lambda$ and remains smooth even below $\Lambda_c$. We have learned from the vertex divergence, however, that we must not trust the result below $T_c$. 
Rescaling the expression to temperature units we recover the exact mean-field result $\chi(T)=\frac{1}{4T}$ for $T>T_c$. 

Although we have seen that all results are consistent with the exact mean-field solution, the present model serves as an important example that reminds us that not always all relevant information is accessible via the magnetic susceptibility, and one must be careful about how to interpret possible flow breakdowns and the absence thereof.


\section{Conclusions}
\label{sec:conclusions}
In this manuscript we have generalized the established pf-FRG flow equations for conventional SU(2)-symmetric spins to SU($N$) symmetry. 
We have demonstrated that the generalization is straight-forward to implement and that it ultimately requires only the adjustment of prefactors in the flow equations. 
The resulting flow equations complement  the generalization to arbitrary spin length $S$ that has recently been formulated  \cite{Baez2017}. 
In principle, both the large-$S$ and the large-$N$ generalizations can even be implemented simultaneously by combining the corresponding prefactors.
 
We have demonstrated that in the $N \to \infty$ limit the solution of the flow equations reproduces the exact mean-field results. 
In combination with the spin-$S$ generalization \cite{Baez2017}, this provides qualitative guidance as to why the pf-FRG approach has proven quantitatively correct in the analysis of many quantum spin models in the past -- despite the fact that the approach  
relies on a number of presumptions and approximations. The flow equations represent what is a delicate balance of diagrams -- those diagrams that reproduce the exact large-$S$ limit help stabilize magnetic order, while those diagrams that dominate in the large-$N$ limit induce spin liquid behavior. While this balance of $1/S$ and $1/N$ diagrams surely does not amount to an entirely unbiased approach, it does explain why the pf-FRG approach has been able to independently capture both magnetically ordered and spin liquid ground states in the phase diagrams of various spin models with competing interactions in the past.

As a case study, we have studied the SU($N$) Heisenberg  model on the square lattice that in the large-$N$ limit is known to host a staggered flux spin liquid. 
We analyzed the symmetry-breaking transition within the pf-FRG framework and found that -- similar to the established flow breakdown associated with a magnetic ordering transition -- the U(1) symmetry breaking can indeed be detected by a breakdown of the smooth susceptibility flow. 
This demonstrates, for the first time, that a flow breakdown in pf-FRG calculations must not always be associated with the transition into a magnetically ordered phase (breaking the conventional SU($N$) spin symmetry) and one should in general be careful about its interpretation. 
In the present example, the staggered flux spin liquid is easy to identify by its featureless real-space correlations. 
However, this does not always need to be the case. To uniquely determine the kind of symmetry-breaking underlying a certain transition one may therefore extend the pf-FRG formalism to explicitly probe the symmetry-broken phase, as demonstrated in our companion manuscript~\onlinecite{Roscher2017}. 

For future developments it would be desirable to not only identify the spontaneous breaking of a symmetry (with or without access to the symmetry-broken phase), but also to directly probe the underlying gauge structure. Doing so would be an important step towards the full characterization of spin liquids within pf-FRG approach. 
This also raises the question whether the pf-FRG approach in its current form can actually capture arbitrary gauge structures: In the present example the underlying U(1) structure is inherently encoded in the definition of the pseudofermions. 
It remains an open issue, however, whether the current pseudofermion approach is equally suited to faithfully capture a $Z_2$ gauge structure. This might be of particular interest for two- and three-dimensional Kitaev models \cite{Kitaev2006,OBrien2016} where the most natural spin decomposition is phrased in terms of Majorana fermions and not complex fermions as discussed here.

Our work also provides a new perspective \cite{Salmhofer2004} on the relevance of the Katanin scheme. It has already been known that the Katanin scheme is essential to overcome what is otherwise a strong bias towards magnetically ordered phases \cite{Reuther2010}. 
Here we report complementary behavior when studying generalized SU($N$) models. The symmetric phase in the large-$N$ limit
 seems easy to reproduce even without the Katanin scheme. But as one moves  into the symmetry-broken phase or away from the mean-field limit and closer to the true SU(2) quantum model the Katanin truncation becomes important and is found to significantly alter the results. 

Last but not least, we identified a situation (in the $N \to \infty$ limit) where a  divergence of the interaction vertex exists, but it is not observable in the magnetic susceptibility. This is a formidable pitfall, which should be circumvented by developing a better understanding of the mechanism behind the flow breakdown and studying signatures thereof not only in the magnetic susceptibility but also in the multitude of  fermionic interaction vertices. 
We leave such a more systematic study to future work.


\begin{acknowledgments}
We thank F. Assaad and J. Reuther for discussions and M. Scherer for joint work on a companion manuscript \cite{Roscher2017}.
This work was partially supported by the DFG within the CRC 1238 (project C03). 
The numerical simulations were performed on the CHEOPS cluster at RRZK Cologne and the JURECA cluster at the Forschungszentrum Juelich.
F.L.B. thanks the Bonn-Cologne Graduate School of Physics and Astronomy (BCGS) for support.
D.R. is supported, in part, by the NSERC of Canada.
\end{acknowledgments}


\bibliography{largeN-realspace}


\appendix
\widetext

\section{SU($N$)-symmetric flow equations}
\begin{figure*}[h]
\centering
\includegraphics[width=0.85\linewidth]{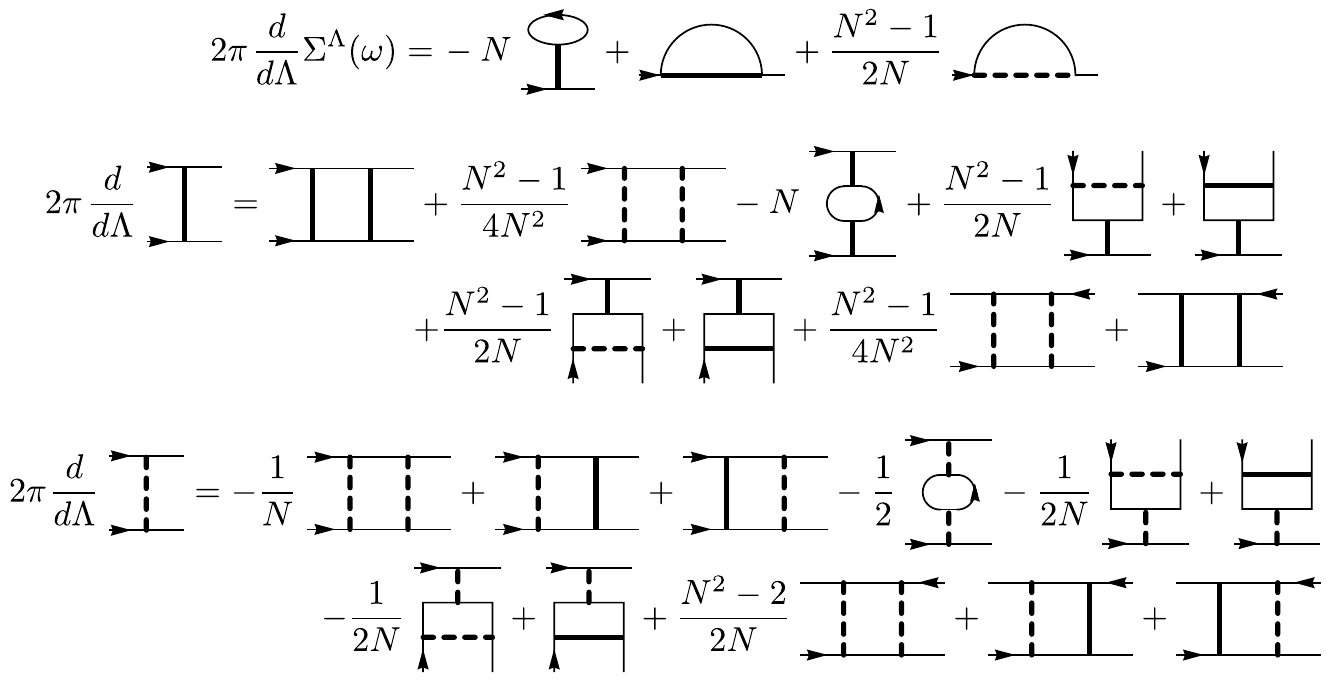}
\caption{{\bf Generalized flow equations} for the SU($N$)-symmetric Heisenberg model. The spin dependency has been calculated explicitly by splitting the two-particle vertex into the two symmetry-allowed terms, a spin contribution (dashed lines) and a density contribution (solid lines). }
\label{fig:flowequationsFull}
\end{figure*}

In this appendix we present the full set of flow equations for generalized $SU($N$)$ Heisenberg models. 
For the sake of brevity we present the flow equations diagrammatically, where the diagrams should be read as 
\begin{equation}
\vcenter{\hbox{\includegraphics[scale=1]{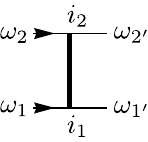}}} = 
\Gamma^\Lambda_{d,i_1i_2}(\omega_{1'}\omega_{2'};\omega_1\omega_2)
\hspace{2cm}
\vcenter{\hbox{\includegraphics[scale=1]{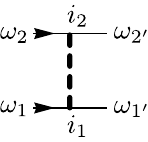}}} = 
\Gamma^\Lambda_{s,i_1i_2}(\omega_{1'}\omega_{2'};\omega_1\omega_2) \,.
\end{equation}
Each term in the two-particle flow equations (Fig.~\ref{fig:flowequationsFull}) has two internal propagator lines, $G_1$ and $G_2$. They should be read as combinations of the full propagator and the single-scale propagator, i.e. the pair of propagators is replaced according to 
\begin{equation}
(G_1,G_2)\to \left( G^\Lambda,S_\mathrm{kat}^\Lambda)+(S_\mathrm{kat}^\Lambda,G^\Lambda\right) 
\end{equation}
where we have explicitly denoted that the Katanin scheme is applied to the single-scale propagator. 
Internal propagator lines in the single-particle flow equation should be understood as single-scale propagators $S^\Lambda(\omega)$ and remain unaffected by the Katanin replacement. 

\end{document}